# Calculated optical properties of BTTzR donor molecule and its derivatives

Giuseppe M. Paternò[1], Andrea Farina[2], Francesco Scotognella[1,3]*

[1] *Center for Nano Science and Technology@PoliMi, Istituto Italiano di Tecnologia, Via Pascoli 70/3, 20133 Milano, Italy*

[2] *Istituto di Fotonica e Nanotecnologie, Consiglio Nazionale delle Ricerche, Piazza Leonardo da Vinci 32, 20133 Milano, Italy*

[3] *Dipartimento di Fisica, Politecnico di Milano, Piazza Leonardo da Vinci 32, 20133 Milano, Italy*

* To whom the correspondence should be addressed: francesco.scotognella@polimi.it

**Abstract**

In this work, we study the light absorption properties of a novel molecule (BTTzR) and its more extended derivatives, which hold promise as electron-donor material in organic solar cells. By employing density functional theory, we observe that the addition of two and three oligothiophene chains to the central benzene ring of the benzo[1,2-b:4,5-b']dithiophene (BDT-T) leads to both a red-shift of the existing peaks and, interestingly, to the development of new blue-shifted features, an effect that can certainly increase the panchromaticity of the molecule in the visible spectral range.

**Introduction**

Together with other materials exhibiting high photovoltaic ability, such as metal-halide perovskites [1–3], semiconducting polymers [4,5], quantum dots [6,7], and carbon nanotubes [8,9], small molecules are widely studied in the solar cell research community. In these regards, atomically precise synthesis of small-molecules allows a negligible batch-to-batch variations [10], which is an important requirement for devices reproducibility and reliability. For instance, fullerene derivatives have represented the workhorse electron-acceptor materials for decades [11–13], together with poly(thiophene) derivatives as electron-donor system [14,15]. However, fullerenes suffer from two important drawbacks, such as a relatively low absorption in the visible spectrum and photo-instabilities (*i.e* dimerization), which limit considerably the power conversion efficiency of the related devices [16].

Therefore, recently many reports have focused on the replacement of fullerene derivatives as electron-acceptor molecule in organic solar cells [17]. For example, newly developed molecules like BTPTT-4F (also called Y6) permit to achieve a noteworthy efficiency of 16.02 % [18–20]. In addition, the research of new donor materials has been fruitful in the last years. In 2020, a novel small molecule donor, BTTzR, has been developed and an all-small-molecule solar cell employing BTTzR as donor shows a power conversion efficiency (PCE) of 13.9% [21].

Given the high interest in these promising function materials, in this work we analyse the light absorption features of BTTzR and its extended derivatives by means of density functional theory. By increasing the length of the thiophene pendants attached to the benzo[1,2-b:4,5-b']dithiophene (BDT-T) central unit of the molecule, we can observe a red-shift of the absorption together with the occurrence of new absorption features at higher energy. This aspect can be important to attain panchromaticity, and thus reach even higher power conversion efficiencies with these promising molecules.

**Methods**

The molecules have been designed with the Avogadro package [22]. We have optimized the geometries and calculated the electronic transitions, by means of the density functional theory, with the ORCA package, developed by Frank Neese and coworkers [23]. In these calculations we have used the B3LYP functional [24]. Moreover, we have employed the Ahlrichs split valence basis set [25] together with the all-electron nonrelativistic basis set SVPalls1 [26,27]. Finally, for thisese calculations we have used the Libint library [28] and the Libxc library [29,30]. The calculations have been performed on a Dell Precision Tower Workstation 7910, which mounts Dual Intel Xeon processors (10 cores, 2.35GHz) and 64 Gb of RAM memory.

**Results and Discussion**

In Figure 1a, we show the optimized geometry of the BTTzR molecule. The dihedral angle between the thiazolo-[5,4-d]thiazole (TTz) group and the BDT-T group along the backbone of the molecule is 14° on one side and 3° on the other side. The dihedral angle between the thiophene molecule and BDT-T is 52° on one side and 55° on the other side. These angles are in good agreement with the ones of optimized geometries reported in the literature [21].

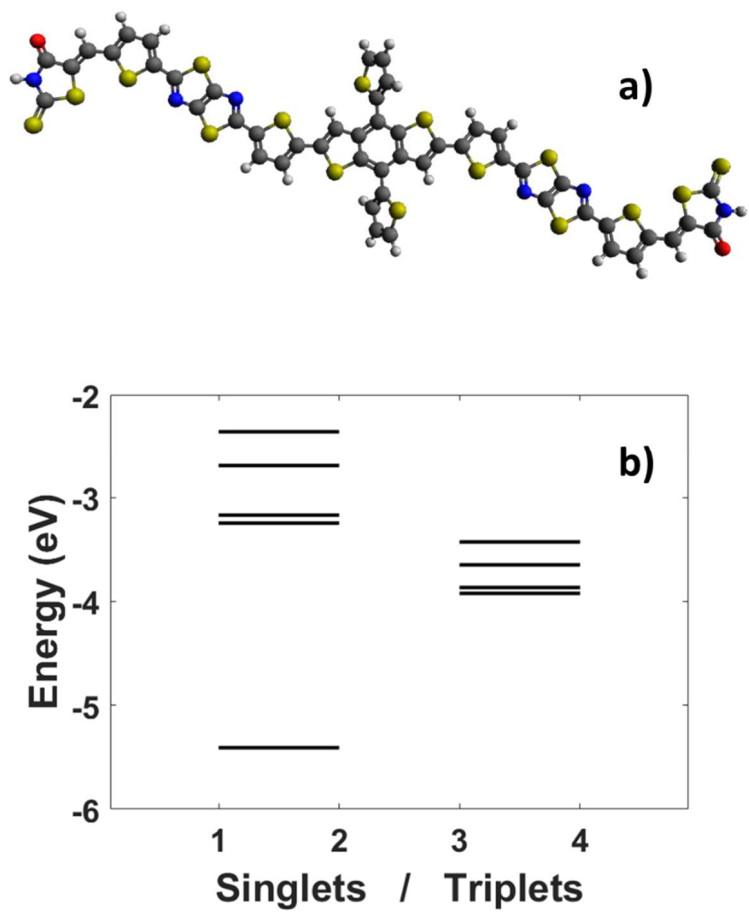

**Figure 1**. a) Optimized geometry of BTTzR molecule; b) energy levels of BTTzR with the lowest four singlet excited states and the lowest four triplet excited states.

In Figure 1b, we show the energy levels of BTTzR, highlighting the ground state at -5.41 eV and the four lowest singlet excited states at -3.25 eV, -3.17 eV, -2.69 eV, and -2.36 eV, respectively. We also show the four lowest triplet excited states at -3.92 eV, -3,87 eV, -3.65 eV, and – 3.43 eV, respectively. In Figure 2a, we show the optimized geometry of BTTzRv2 (where _v2 is used to simply highlight that it is a second derivative/version of the molecule) with two thiophene molecules attached to the BDT-T unit.

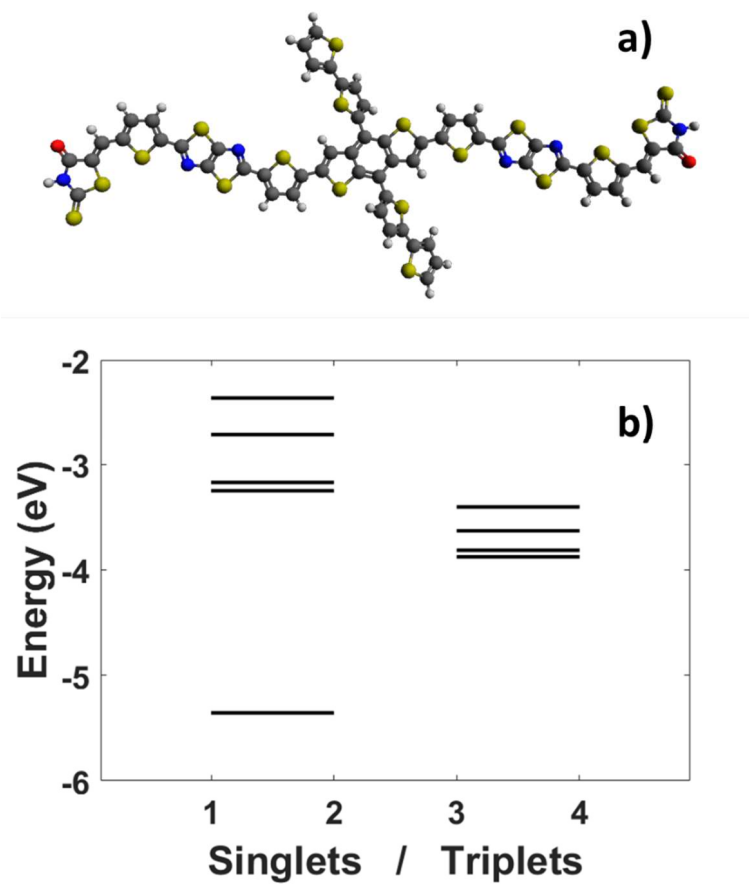

**Figure 2**. a) Optimized geometry of BTTzRv2 molecule, with two thiophenes attached to the BDT-T unit; b) energy levels of BTTzRv2 with the lowest four singlet excited states and the lowest four triplet excited states.

In Figure 2b, we show the energy levels of BTTRv2. The ground state is at -5.36 eV and the lowest four singlet excited states are at -3.25 eV, -3.17 eV, -2.71 eV, and -2.36 eV, respectively. With respect to BTTzR, we notice a ground state higher in energy. The lowest four triplet excited states are at -3.88 eV, -3,81 eV, -3.63 eV, and – 3.40 eV, respectively.

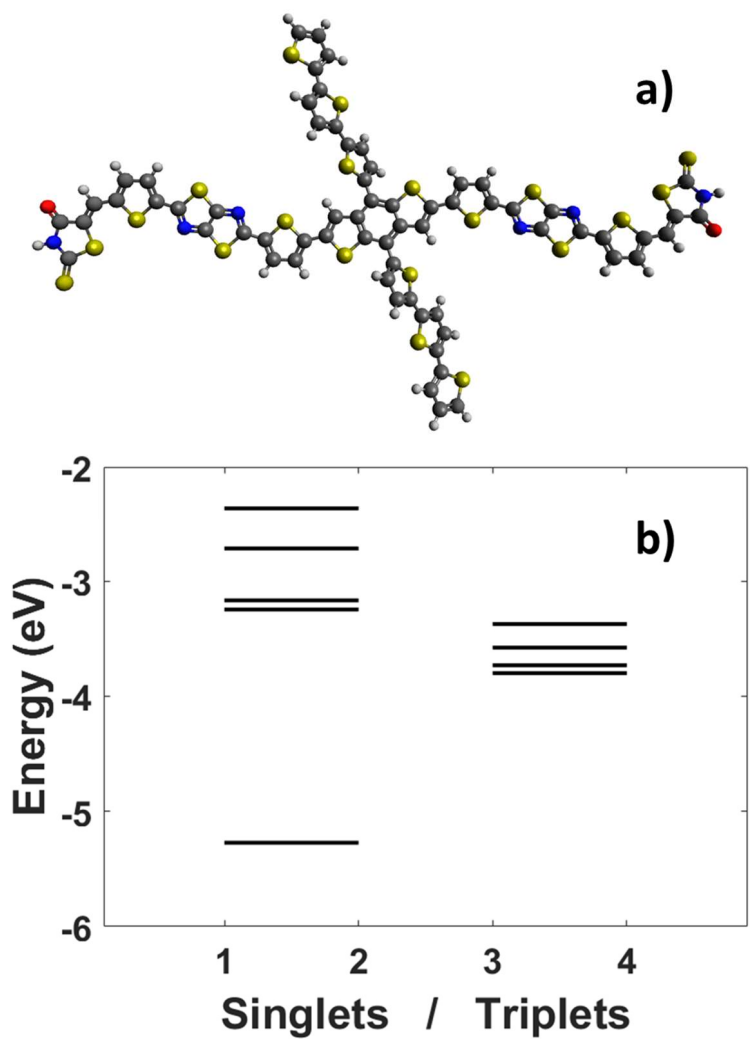

**Figure 3**. a) Optimized geometry of BTTzRv3 molecule, with three thiophenes attached to the BDT-T unit; b) energy levels of BTTzRv3 with the lowest four singlet excited states and the lowest four triplet excited states.

In Figure 3a, we show the optimized geometry of BTTzRv3 in which three thiophene molecules are attached to the BDT-T unit. In Figure 3b the energy levels of the molecule are depicted with the ground state at -5.27 eV (higher with respect of BTTzR and BTTzRv2), the lowest four singlet excited states at -3.23 eV, -3.17 eV, -2.71 eV, and -2.36 eV, respectively, the lowest four triplet excited states at -3.80 eV, -3.73 eV, -3.58 eV, and -3.37 eV, respectively.

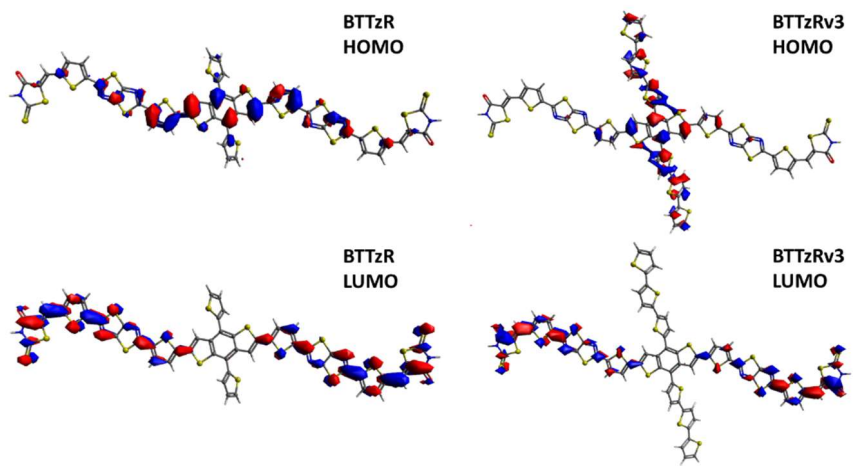

**Figure 4.** Highest occupied molecular orbitals (HOMOs) and lowest unoccupied molecular orbitals (LUMOs) of BTTzR and BTTzRv3.

In Figure 4 we show the highest occupied molecular orbitals (HOMOs) and lowest unoccupied molecular orbitals (LUMOs) of BTTzR and BTTzRv3, highlighting that such orbitals, in particular the HOMO orbital, are significantly delocalized over the oligothiophene pendants in BTTzRv3.

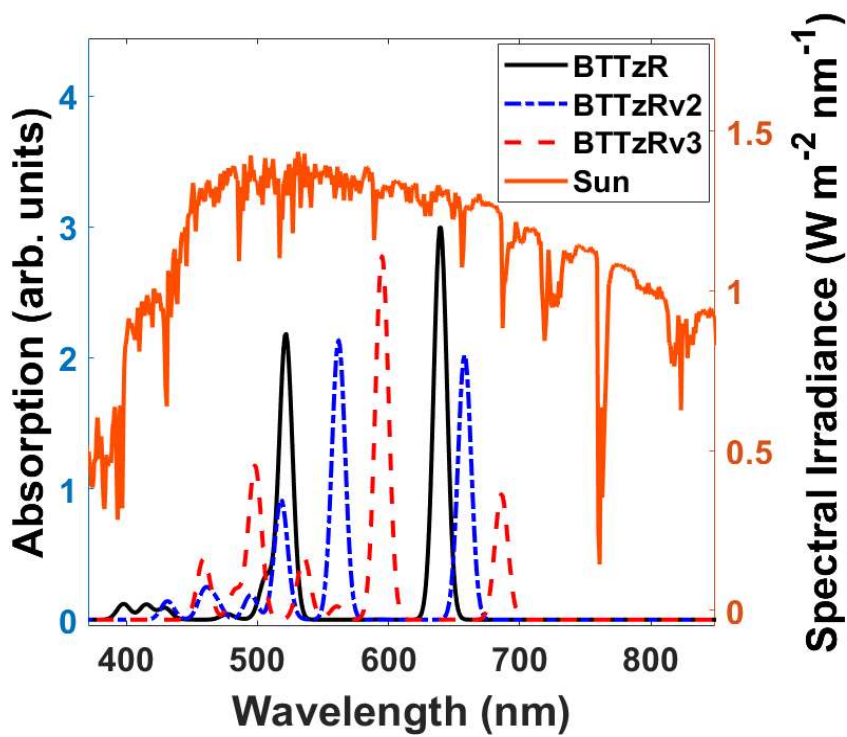

**Figure 5**. Absorption spectra in the visible spectral range of BTTzR (solid black curve), of BTTzRv2 (dotted dashed blue curve) and of BTTzRv3 (dashed red curve). In dark orange the direct and circumsolar irradiation of the Sun [31].

In Figure 5, we show the absorption spectra of BTTzR (solid black curve), of BTTzRv2 (dotted dashed blue curve) and of BTTzRv3 (dashed red curve) in the visible spectral range. The calculated lowest electronic transition is at lower energy with respect to the experimental one (in solution it has been reported a maximum absorption peak at 528 nm) [21] and this could be due to a discrepancy between the dihedral angles in the calculated optimized geometry and the actual dihedral angles of the molecules in solution. In the Supporting Information we report the lowest 16 electronic (singlet to singlet) transitions and the lowest 16 triplets. Here, we notice a red shift of the lowest transition from BTTzR to BTTzRv2 and BTTzRv3, due to an increased delocalization thanks to the longest thiophene chains attached to the BDT-T unit. Moreover, with BTTzR we observe two intense absorption peaks and with BTTzRv2 three intense absorption peaks. Finally, with BTTzRv3 five intense absorption peaks at 685, 595, 535, 498, and 459 nm, respectively, that span a significant range of the visible spectrum. The larger number of absorption peaks of BTTzRv3 can lead to a potential increased absorption of the Sun irradiation (the orange curve in Figure 4 displays the direct and circumsolar irradiation of the Sun, reference Air Mass 1.5 [31]), useful for the fabrication of efficient photovoltaic cells. We also calculate the cation of BTTzRv3 and we report in the Supporting Information the optimized geometry and lowest 16 electronic transitions. It is remarkable a strong red shift of the lowest electronic transitions in the cation with respect to the ones of the neutral molecules with transitions at about 2.4 micrometers with a non-negligible oscillator strength.

## Conclusion

In this work, we have studied the optical properties in the visible range of the donor molecule BTTzR and of two derivatives of BTTzR in which we have increased the number of thiophene molecules attached to the BDT-T central unit. We have calculated the singlet and triplet excited states noticing a red-shift of such levels by increasing the number of thiophene molecules. Furthermore, addition of the thiophene pendants leads to the appearance of new blue-shifted absorption features, which could lead to an increased panchromaticity and power conversion efficiencies in the related solar cells.

## Acknowledgement

G.M.P. and F.S. thanks Fondazione Cariplo (grant No. 2018-0979) for financial support. This project has received funding from the European Research Council (ERC) under the European Union's Horizon 2020 research and innovation programme (grant agreement No. [816313]).

## References

[1]  J.M. Ball, A. Petrozza, Defects in perovskite-halides and their effects in solar cells, Nature Energy. 1 (2016) 1–13. https://doi.org/10.1038/nenergy.2016.149.
[2]  O.M. Bakr, M. Leite, N. Pradhan, Energy Spotlight, ACS Energy Lett. 5 (2020) 1328–1329. https://doi.org/10.1021/acsenergylett.0c00687.
[3]  J. Yuan, A. Hazarika, Q. Zhao, X. Ling, T. Moot, W. Ma, J.M. Luther, Metal Halide Perovskites in Quantum Dot Solar Cells: Progress and Prospects, Joule. 4 (2020) 1160–1185. https://doi.org/10.1016/j.joule.2020.04.006.

[4] I. Etxebarria, J. Ajuria, R. Pacios, Solution-processable polymeric solar cells: A review on materials, strategies and cell architectures to overcome 10%, Organic Electronics. 19 (2015) 34–60. https://doi.org/10.1016/j.orgel.2015.01.014.

[5] H. Sun, X. Guo, A. Facchetti, High-Performance n-Type Polymer Semiconductors: Applications, Recent Development, and Challenges, Chem. 6 (2020) 1310–1326. https://doi.org/10.1016/j.chempr.2020.05.012.

[6] O. Ouellette, A. Lesage-Landry, B. Scheffel, S. Hoogland, F.P.G. de Arquer, E.H. Sargent, Spatial Collection in Colloidal Quantum Dot Solar Cells, Advanced Functional Materials. 30 (2020) 1908200. https://doi.org/10.1002/adfm.201908200.

[7] H. Lee, H.-J. Song, M. Shim, C. Lee, Towards the commercialization of colloidal quantum dot solar cells: perspectives on device structures and manufacturing, Energy Environ. Sci. 13 (2020) 404–431. https://doi.org/10.1039/C9EE03348C.

[8] A.J. Ferguson, J.L. Blackburn, N. Kopidakis, Fullerenes and carbon nanotubes as acceptor materials in organic photovoltaics, Materials Letters. 90 (2013) 115–125. https://doi.org/10.1016/j.matlet.2012.08.145.

[9] G. Soavi, F. Scotognella, D. Viola, T. Hefner, T. Hertel, G. Cerullo, G. Lanzani, High energetic excitons in carbon nanotubes directly probe charge-carriers, Sci Rep. 5 (2015) 1–5. https://doi.org/10.1038/srep09681.

[10] H. Bin, I. Angunawela, B. Qiu, F.J.M. Colberts, M. Li, M.J. Dyson, M.M. Wienk, H. Ade, Y. Li, R.A.J. Janssen, Precise Control of Phase Separation Enables 12% Efficiency in All Small Molecule Solar Cells, Advanced Energy Materials. 10 (2020) 2001589. https://doi.org/10.1002/aenm.202001589.

[11] G. Paternò, A. J. Warren, J. Spencer, G. Evans, V. García Sakai, J. Blumberger, F. Cacialli, Micro-focused X-ray diffraction characterization of high-quality [6,6]-phenyl-C 61 -butyric acid methyl ester single crystals without solvent impurities, Journal of Materials Chemistry C. 1 (2013) 5619–5623. https://doi.org/10.1039/C3TC31075B.

[12] G. Tregnago, M. Wykes, G.M. Paternò, D. Beljonne, F. Cacialli, Low-Temperature Photoluminescence Spectroscopy of Solvent-Free PCBM Single-Crystals, J. Phys. Chem. C. 119 (2015) 11846–11851. https://doi.org/10.1021/acs.jpcc.5b02345.

[13] G.M. Lazzerini, G.M. Paternò, G. Tregnago, N. Treat, N. Stingelin, A. Yacoot, F. Cacialli, Traceable atomic force microscopy of high-quality solvent-free crystals of [6,6]-phenyl-C61-butyric acid methyl ester, Appl. Phys. Lett. 108 (2016) 053303. https://doi.org/10.1063/1.4941227.

[14] G.M. Paternò, M.W.A. Skoda, R. Dalgliesh, F. Cacialli, V.G. Sakai, Tuning Fullerene Intercalation in a Poly (thiophene) derivative by Controlling the Polymer Degree of Self-Organisation, Scientific Reports. 6 (2016) 34609. https://doi.org/10.1038/srep34609.

[15] G.M. Paternò, J.R. Stewart, A. Wildes, F. Cacialli, V.G. Sakai, Neutron polarisation analysis of Polymer:Fullerene blends for organic photovoltaics, Polymer. 105 (2016) 407–413. https://doi.org/10.1016/j.polymer.2016.07.079.

[16] C. Yan, S. Barlow, Z. Wang, H. Yan, A.K.-Y. Jen, S.R. Marder, X. Zhan, Non-fullerene acceptors for organic solar cells, Nat Rev Mater. 3 (2018) 18003. https://doi.org/10.1038/natrevmats.2018.3.

[17] R.S. Gurney, D.G. Lidzey, T. Wang, A review of non-fullerene polymer solar cells: from device physics to morphology control, Rep. Prog. Phys. 82 (2019) 036601. https://doi.org/10.1088/1361-6633/ab0530.

[18] J. Yuan, Y. Zhang, L. Zhou, G. Zhang, H.-L. Yip, T.-K. Lau, X. Lu, C. Zhu, H. Peng, P.A. Johnson, M. Leclerc, Y. Cao, J. Ulanski, Y. Li, Y. Zou, Single-Junction Organic Solar Cell with over 15%

Efficiency Using Fused-Ring Acceptor with Electron-Deficient Core, Joule. 3 (2019) 1140–1151. https://doi.org/10.1016/j.joule.2019.01.004.

[19]    B. Fan, D. Zhang, M. Li, W. Zhong, Z. Zeng, L. Ying, F. Huang, Y. Cao, Achieving over 16% efficiency for single-junction organic solar cells, Sci. China Chem. 62 (2019) 746–752. https://doi.org/10.1007/s11426-019-9457-5.

[20]    A. Farina, G.M. Paternò, F. Scotognella, Optical properties of recent non-fullerene molecular acceptors for bulk heterojunction solar cells, Results in Physics. 19 (2020) 103633. https://doi.org/10.1016/j.rinp.2020.103633.

[21]    Y. Wang, Y. Wang, L. Zhu, H. Liu, J. Fang, X. Guo, F. Liu, Z. Tang, M. Zhang, Y. Li, A novel wide-bandgap small molecule donor for high efficiency all-small-molecule organic solar cells with small non-radiative energy losses, Energy Environ. Sci. 13 (2020) 1309–1317. https://doi.org/10.1039/C9EE04199K.

[22]    M.D. Hanwell, D.E. Curtis, D.C. Lonie, T. Vandermeersch, E. Zurek, G.R. Hutchison, Avogadro: an advanced semantic chemical editor, visualization, and analysis platform, J Cheminform. 4 (2012) 17. https://doi.org/10.1186/1758-2946-4-17.

[23]    F. Neese, The ORCA program system, WIREs Comput Mol Sci. 2 (2012) 73–78. https://doi.org/10.1002/wcms.81.

[24]    C. Lee, W. Yang, R.G. Parr, Development of the Colle-Salvetti correlation-energy formula into a functional of the electron density, Physical Review B. 37 (1988) 785–789. https://doi.org/10.1103/PhysRevB.37.785.

[25]    A. Schäfer, H. Horn, R. Ahlrichs, Fully optimized contracted Gaussian basis sets for atoms Li to Kr, The Journal of Chemical Physics. 97 (1992) 2571–2577. https://doi.org/10.1063/1.463096.

[26]    A. Schäfer, C. Huber, R. Ahlrichs, Fully optimized contracted Gaussian basis sets of triple zeta valence quality for atoms Li to Kr, The Journal of Chemical Physics. 100 (1994) 5829–5835. https://doi.org/10.1063/1.467146.

[27]    K. Eichkorn, F. Weigend, O. Treutler, R. Ahlrichs, Auxiliary basis sets for main row atoms and transition metals and their use to approximate Coulomb potentials, Theoretical Chemistry Accounts: Theory, Computation, and Modeling (Theoretica Chimica Acta). 97 (1997) 119–124. https://doi.org/10.1007/s002140050244.

[28]    E.~F.~Valeev, A library for the evaluation of molecular integrals of many-body operators over Gaussian functions, (2014). http://libint.valeyev.net/.

[29]    S. Lehtola, C. Steigemann, M.J.T. Oliveira, M.A.L. Marques, Recent developments in libxc — A comprehensive library of functionals for density functional theory, SoftwareX. 7 (2018) 1–5. https://doi.org/10.1016/j.softx.2017.11.002.

[30]    M.A.L. Marques, M.J.T. Oliveira, T. Burnus, Libxc: A library of exchange and correlation functionals for density functional theory, Computer Physics Communications. 183 (2012) 2272–2281. https://doi.org/10.1016/j.cpc.2012.05.007.

[31]    Reference Air Mass 1.5 Spectra, (n.d.). https://www.nrel.gov/grid/solar-resource/spectra-am1.5.html (accessed January 30, 2021).

# Supporting Information

Electronic transitions of BTTzR

```
--------------------------------------------------------------------------------
State   Energy    Wavelength   fosc          P2         PX        PY        PZ
        (cm-1)    (nm)                       (au**2)    (au)      (au)      (au)
--------------------------------------------------------------------------------
   1    15636.7     639.5      1.477920350   0.15794    0.39145  -0.06810   0.00855
   2    16841.4     593.8      0.000148771   0.00002   -0.00239   0.00046  -0.00335
   3    19169.3     521.7      1.118158338   0.14649   -0.38074   0.03903  -0.00286
   4    19586.7     510.5      0.003746041   0.00050    0.01747  -0.00375  -0.01350
   5    19745.5     506.4      0.189039986   0.02551    0.15151  -0.05018   0.00618
   6    20479.1     488.3      0.001839254   0.00026   -0.00895   0.00372   0.01278
   7    20837.2     479.9      0.000326065   0.00005   -0.00000   0.00111   0.00672
   8    20847.0     479.7      0.000323086   0.00005    0.00010  -0.00151   0.00661
   9    20968.3     476.9      0.015320480   0.00220   -0.00449  -0.04615   0.00672
  10    22540.7     443.6      0.000874983   0.00013    0.00050   0.00056   0.01159
  11    22775.7     439.1      0.000153792   0.00002    0.00158   0.00102   0.00452
  12    23280.6     429.5      0.028952128   0.00461   -0.06694  -0.00978   0.00543
  13    24070.7     415.4      0.012709640   0.00209   -0.03382   0.03077   0.00059
  14    24155.1     414.0      0.002060287   0.00034   -0.01378   0.01192  -0.00286
  15    25124.1     398.0      0.061174498   0.01050   -0.09766   0.03110  -0.00055
  16    25333.4     394.7      0.000003101   0.00000    0.00021   0.00011  -0.00069
  17    11981.2     834.6      spin forbidden (mult=3)
  18    12439.2     803.9      spin forbidden (mult=3)
  19    14194.1     704.5      spin forbidden (mult=3)
  20    15975.1     626.0      spin forbidden (mult=3)
  21    17394.4     574.9      spin forbidden (mult=3)
  22    17571.9     569.1      spin forbidden (mult=3)
  23    17997.6     555.6      spin forbidden (mult=3)
  24    19105.0     523.4      spin forbidden (mult=3)
  25    19111.4     523.2      spin forbidden (mult=3)
  26    19420.1     514.9      spin forbidden (mult=3)
  27    19885.6     502.9      spin forbidden (mult=3)
  28    20426.2     489.6      spin forbidden (mult=3)
  29    21344.0     468.5      spin forbidden (mult=3)
  30    21648.1     461.9      spin forbidden (mult=3)
  31    22325.9     447.9      spin forbidden (mult=3)
  32    22379.8     446.8      spin forbidden (mult=3)
```

Electronic transitions of BTTzRv2

```
--------------------------------------------------------------------------------
State   Energy    Wavelength   fosc          T2         TX        TY        TZ
        (cm-1)    (nm)                       (au**2)    (au)      (au)      (au)
--------------------------------------------------------------------------------
   1    15197.0     658.0      2.021711980   43.79628   6.61637   0.01638   0.14028
   2    16215.0     616.7      0.001142288    0.02319  -0.03198  -0.01493  -0.14814
   3    17796.4     561.9      2.135623480   39.50656   6.20020  -1.03139   0.01954
   4    18888.6     529.4      0.009514354    0.16583   0.40637  -0.00009  -0.02627
   5    19294.0     518.3      0.585182800    9.98491  -3.09206  -0.58201  -0.29207
   6    19317.0     517.7      0.334647470    5.70326   2.34523   0.43517  -0.11751
   7    20202.4     495.0      0.188798469    3.07661   1.39238  -1.06088  -0.11144
   8    20598.5     485.5      0.001422198    0.02273  -0.03882   0.03251   0.14201
   9    20833.9     480.0      0.000015203    0.00024  -0.00021   0.00232   0.01532
  10    20844.4     479.7      0.000017564    0.00028   0.00113   0.00339  -0.01627
  11    21335.9     468.7      0.127834201    1.97248   1.32198  -0.47219  -0.04352
  12    21749.0     459.8      0.216770939    3.28123  -1.64249   0.76154   0.05920
  13    21979.1     455.0      0.000775872    0.01162   0.03148  -0.01668  -0.10174
  14    22418.7     446.1      0.001309972    0.01924  -0.12708   0.03627   0.04209
```

| State | Energy (cm-1) | Wavelength (nm) | fosc | T2 (au**2) | TX (au) | TY (au) | TZ (au) |
|---|---|---|---|---|---|---|---|
| 15 | 23175.2 | 431.5 | 0.141339165 | 2.00778 | 1.41282 | -0.10821 | 0.00360 |
| 16 | 23576.2 | 424.2 | 0.000177383 | 0.00248 | 0.02972 | -0.03902 | -0.00843 |
| 17 | 11928.1 | 838.4 | spin forbidden (mult=3) | | | | |
| 18 | 12436.2 | 804.1 | spin forbidden (mult=3) | | | | |
| 19 | 13927.3 | 718.0 | spin forbidden (mult=3) | | | | |
| 20 | 15751.1 | 634.9 | spin forbidden (mult=3) | | | | |
| 21 | 16578.2 | 603.2 | spin forbidden (mult=3) | | | | |
| 22 | 16973.7 | 589.1 | spin forbidden (mult=3) | | | | |
| 23 | 17457.4 | 572.8 | spin forbidden (mult=3) | | | | |
| 24 | 18732.8 | 533.8 | spin forbidden (mult=3) | | | | |
| 25 | 19076.1 | 524.2 | spin forbidden (mult=3) | | | | |
| 26 | 19101.7 | 523.5 | spin forbidden (mult=3) | | | | |
| 27 | 19108.0 | 523.3 | spin forbidden (mult=3) | | | | |
| 28 | 19633.6 | 509.3 | spin forbidden (mult=3) | | | | |
| 29 | 19730.0 | 506.8 | spin forbidden (mult=3) | | | | |
| 30 | 20225.5 | 494.4 | spin forbidden (mult=3) | | | | |
| 31 | 20759.1 | 481.7 | spin forbidden (mult=3) | | | | |
| 32 | 20785.8 | 481.1 | spin forbidden (mult=3) | | | | |

Electronic transitions of BTTzRv3

| State | Energy (cm-1) | Wavelength (nm) | fosc | T2 (au**2) | TX (au) | TY (au) | TZ (au) |
|---|---|---|---|---|---|---|---|
| 1 | 14575.0 | 686.1 | 0.958641369 | 21.65330 | 4.61040 | 0.59038 | 0.22124 |
| 2 | 15456.2 | 647.0 | 0.000546135 | 0.01163 | -0.00565 | -0.01865 | -0.10608 |
| 3 | 16801.2 | 595.2 | 2.689394058 | 52.69738 | -7.17553 | 1.09931 | -0.02526 |
| 4 | 16927.2 | 590.8 | 0.128157372 | 2.49249 | -1.56308 | 0.21671 | 0.04793 |
| 5 | 17834.4 | 560.7 | 0.101340088 | 1.87068 | 1.36645 | -0.00568 | 0.05893 |
| 6 | 18115.6 | 552.0 | 0.001088340 | 0.01978 | -0.05628 | -0.01278 | 0.12825 |
| 7 | 18677.8 | 535.4 | 0.474955273 | 8.37148 | -2.05719 | -1.99386 | -0.40489 |
| 8 | 19566.3 | 511.1 | 0.003204190 | 0.05391 | 0.01187 | -0.01572 | -0.23135 |
| 9 | 20071.7 | 498.2 | 1.175644149 | 19.28263 | -4.23088 | 1.17007 | 0.11496 |
| 10 | 20693.8 | 483.2 | 0.221034971 | 3.51639 | 1.80677 | -0.50116 | -0.02824 |
| 11 | 20834.8 | 480.0 | 0.000026763 | 0.00042 | -0.01180 | 0.00536 | 0.01597 |
| 12 | 20846.8 | 479.7 | 0.000018053 | 0.00029 | 0.00355 | 0.00334 | -0.01616 |
| 13 | 20877.7 | 479.0 | 0.002634004 | 0.04153 | -0.20215 | 0.02540 | -0.00500 |
| 14 | 21370.5 | 467.9 | 0.000270139 | 0.00416 | -0.03430 | -0.00684 | -0.05421 |
| 15 | 21783.9 | 459.1 | 0.463656807 | 7.00706 | -1.76922 | 1.93890 | 0.34289 |
| 16 | 21869.8 | 457.3 | 0.000468562 | 0.00705 | 0.02898 | -0.05660 | 0.05486 |
| 17 | 11880.7 | 841.7 | spin forbidden (mult=3) | | | | |
| 18 | 12430.2 | 804.5 | spin forbidden (mult=3) | | | | |
| 19 | 13670.3 | 731.5 | spin forbidden (mult=3) | | | | |
| 20 | 15318.3 | 652.8 | spin forbidden (mult=3) | | | | |
| 21 | 15658.5 | 638.6 | spin forbidden (mult=3) | | | | |
| 22 | 16299.8 | 613.5 | spin forbidden (mult=3) | | | | |
| 23 | 16330.6 | 612.3 | spin forbidden (mult=3) | | | | |
| 24 | 16625.4 | 601.5 | spin forbidden (mult=3) | | | | |
| 25 | 17285.8 | 578.5 | spin forbidden (mult=3) | | | | |
| 26 | 17689.6 | 565.3 | spin forbidden (mult=3) | | | | |
| 27 | 17861.0 | 559.9 | spin forbidden (mult=3) | | | | |
| 28 | 18271.8 | 547.3 | spin forbidden (mult=3) | | | | |
| 29 | 18935.2 | 528.1 | spin forbidden (mult=3) | | | | |
| 30 | 19102.5 | 523.5 | spin forbidden (mult=3) | | | | |
| 31 | 19110.3 | 523.3 | spin forbidden (mult=3) | | | | |
| 32 | 19804.3 | 504.9 | spin forbidden (mult=3) | | | | |

Electronic transitions of BTTzRv3 (cation)

```
------------------------------------------------------------------------------
         ABSORPTION SPECTRUM VIA TRANSITION ELECTRIC DIPOLE MOMENTS
------------------------------------------------------------------------------
State   Energy    Wavelength   fosc          T2         TX        TY        TZ
        (cm-1)      (nm)                  (au**2)      (au)      (au)      (au)
------------------------------------------------------------------------------
    1    3821.8     2616.6    0.000035512   0.00306    0.00215  -0.00750   0.05476
    2    4144.9     2412.6    1.127426796  89.54654   -9.30594   1.71631  -0.01690
    3    5845.9     1710.6    0.858123821  48.32532   -1.74161   6.62170   1.20216
    4    6902.3     1448.8    0.000008216   0.00039    0.00643  -0.00589  -0.01777
    5    6951.6     1438.5    0.000000590   0.00003    0.00505   0.00155   0.00022
    6    6959.8     1436.8    0.000000049   0.00000    0.00088   0.00011  -0.00124
    7    9340.6     1070.6    0.037920617   1.33653    1.08821   0.36698   0.13286
    8    9493.7     1053.3    0.000005839   0.00020   -0.00653  -0.00436   0.01186
    9   12309.8      812.4    0.218494475   5.84339    2.32124   0.64705   0.19121
   10   12502.9      799.8    0.000203631   0.00536    0.07009   0.01850   0.01036
   11   12694.7      787.7    0.000004148   0.00011    0.00968   0.00319   0.00191
   12   13844.5      722.3    0.067808081   1.61243   -1.13311  -0.54087  -0.18961
   13   13967.3      716.0    0.837181344  19.73257    4.11203   1.62253   0.43721
   14   14832.5      674.2    0.038352978   0.85125   -0.88501  -0.25199  -0.06713
   15   14965.4      668.2    0.000000375   0.00001    0.00122   0.00167   0.00199
   16   15221.9      656.9    0.037140601   0.80326    0.88322   0.15177  -0.01211
```

Optimized geometry of BTTzR

```
 C   -1.03432534647874     1.27609255762794    -0.33137107176170
 C    0.27403668772595     0.77206200386489    -0.32629276972730
 C    1.31777848853538     1.73292170895384    -0.39060869321699
 C    1.00622053152298     3.12668827971284    -0.47116984855369
 C   -0.30195332940943     3.63068729017538    -0.47607120928352
 C   -1.34585609179201     2.67034527322190    -0.40406811597165
 C   -2.76155877295242     2.89393172383210    -0.44846623946772
 C   -3.51140419595948     1.74252726969653    -0.40173137889706
 H   -3.20627399697096     3.88616637207674    -0.53686884075873
 S   -2.49644072905349     0.30175611306365    -0.29926431235524
 S    2.46758108074666     4.10197970906161    -0.49496633341710
 C    2.73302654190113     1.51075521604481    -0.33547252236394
 C    3.48345012070674     2.66207001601703    -0.38342137694998
 H    3.17705991716617     0.51749967787099    -0.25420438183767
 C   -4.95204754165998     1.60660984924200    -0.42456134161912
 C    4.92313577455641     2.79856570604829    -0.34354771903901
 C   -5.70365715791396     0.45699354546962    -0.63713907813342
 C   -7.09714278985245     0.68337992133148    -0.59505933599595
 C   -7.42701858854423     2.00813652061200    -0.34670086605622
 C    5.67780001680938     3.96240188941767    -0.43537181201051
 C    7.06933535988508     3.73657318036829    -0.35483545430375
 C    7.39602867481209     2.39702572510061    -0.19951840972517
 S   -5.99196137713501     2.98567797899479    -0.15729876747517
 S    5.95988720181063     1.40369570730743    -0.15335388798506
 H   -5.25469930017819    -0.51896170422408    -0.83425529167658
 H   -7.84224983595389    -0.10144378958953    -0.74627086759200
 H    5.23277342109370     4.95203254658343    -0.56047862391774
 H    7.81532135823089     4.53317424100294    -0.40813363153130
 C   -8.72459562279432     2.62437643850855    -0.22324737405611
 N   -8.90763133930003     3.90449034257623     0.02647756146619
 S  -10.21892986024576     1.66167012264567    -0.39536478412269
 C  -11.12327972986427     3.12305975831086    -0.10115527228125
 C  -10.22232307401602     4.17629620308748     0.09351926910670
 S  -11.12505763406678     5.63571973110946     0.39301652715985
 N  -12.43661882923870     3.39563676539794    -0.02807434806583
 C  -12.61911545985295     4.67414703031693     0.22777494451005
 C    8.69059424829032     1.77564418354143    -0.07299033600227
 S   10.18525541601101     2.75271573605439    -0.10440491803758
 N    8.87067693574141     0.47952393264343     0.07724297547644
 C   11.08523508713174     1.27402075943321     0.10392547361683
 C   10.18313280396382     0.20615302525006     0.17438855640079
 N   12.39605704820200     0.99945412597992     0.20694076188693
 S   11.08158744626562    -1.27098368267780     0.38810735925456
 C   12.57563450255396    -0.29550207928295     0.36301266985834
 C   13.87126473401632    -0.91131077006059     0.51281595639647
 C   14.19834703009136    -2.25051794177097     0.69989974190636
 C   15.58416475867814    -2.46619948051028     0.81669061680675
 C   16.34076167764582    -1.29770331377516     0.72021301420771
 S   15.29855445095548     0.08931823909396     0.47911564867201
 H   13.45160998910886    -3.04619619058440     0.75332499090174
 H   16.04044609645411    -3.44668924891803     0.97205550244181
 C  -13.91794153400551     5.28385765956159     0.37530363962224
 C  -14.24730839080022     6.60427279468580     0.66409611510350
 C  -15.63605698234199     6.82047062446073     0.73691741177195
 C  -16.39242886044340     5.67098209683219     0.50498503462757
 S  -15.34642843071443     4.30148241102280     0.19148532598585
 H  -16.09460365697263     7.78745009509370     0.95663689606141
```

| | | | |
|---|---|---|---|
| H | -13.50023983064748 | 7.38588856279933 | 0.82077817326793 |
| C | -17.82358698086899 | 5.61012361549181 | 0.52343316269248 |
| C | -18.66870710109356 | 4.56235581538682 | 0.32088983991282 |
| H | -18.33149436700356 | 6.55864251805709 | 0.73764825038354 |
| S | -18.28836377208767 | 2.87619699987120 | -0.04449691217299 |
| C | -20.00990842829912 | 2.41568379435028 | -0.09389620304844 |
| S | -20.55696123194333 | 0.90312950857850 | -0.39673961758567 |
| N | -20.76940926611964 | 3.52827133420867 | 0.16084374781568 |
| C | -20.14261806210255 | 4.75667369417538 | 0.40078701496481 |
| H | -21.78445481838444 | 3.45853588541995 | 0.17534795909875 |
| O | -20.73500944499808 | 5.78668309472330 | 0.63246864965167 |
| C | 17.76890339977175 | -1.23186836574358 | 0.81111761805869 |
| C | 18.61312447322857 | -0.16661006338730 | 0.73611758612024 |
| H | 18.27459947621905 | -2.19222605468352 | 0.97074148551617 |
| S | 18.23467721572262 | 1.54142865813957 | 0.49080408776586 |
| C | 19.95169492140755 | 2.01456145671918 | 0.56522693514973 |
| C | 20.08318396951440 | -0.35842665137544 | 0.87239267687575 |
| O | 20.67365666058691 | -1.40048072190211 | 1.04828623517068 |
| N | 20.70883958482114 | 0.88920798465686 | 0.76525552187636 |
| H | 21.72100881046677 | 0.96374305590862 | 0.83815838530518 |
| S | 20.49769857961514 | 3.55027498839132 | 0.41568950013680 |
| C | -0.55707713595469 | 5.08098947260031 | -0.55472090340624 |
| C | -0.06079555694491 | 5.98631278898650 | -1.47165440579823 |
| C | -0.47659816738843 | 7.32888956638763 | -1.22775129808863 |
| C | -1.28128377495585 | 7.43888638431072 | -0.12150309715466 |
| H | -0.19214978583189 | 8.17777069754043 | -1.85393449434482 |
| H | 0.57133058584566 | 5.68779999733573 | -2.31082488920444 |
| S | -1.53917522391612 | 5.90871760674916 | 0.63729660272010 |
| H | -1.73861717509467 | 8.33758097275915 | 0.29384866319164 |
| C | 0.52934894608372 | -0.67831790665735 | -0.25112616965809 |
| C | 0.02856999583332 | -1.58734165937175 | 0.65971780741347 |
| C | 0.44652686403234 | -2.92874483590331 | 0.41328769746377 |
| C | 1.25825100128504 | -3.03428375049349 | -0.68822083049483 |
| H | 0.15884458091329 | -3.77993034228683 | 1.03485172265008 |
| H | -0.60862120570034 | -1.29239295072599 | 1.49631290877042 |
| S | 1.51920524375908 | -1.50129245272440 | -1.44033880075326 |
| H | 1.71877807812697 | -3.93114659520399 | -1.10400138931541 |

Optimized geometry of BTTzRv2

| | | | |
|---|---|---|---|
| C | -1.07032166562176 | 1.35264620979801 | -0.41527214705539 |
| C | 0.23502248434779 | 0.83850547079051 | -0.44406349312342 |
| C | 1.28572235736253 | 1.79534924750544 | -0.47837970285368 |
| C | 0.98381406657914 | 3.19323147360716 | -0.49990919589501 |
| C | -0.32122016307571 | 3.70724818959473 | -0.47092811960949 |
| C | -1.37183689637146 | 2.75070805141579 | -0.42849934050118 |
| C | -2.78599317147635 | 2.98589466001414 | -0.45888558047619 |
| C | -3.54429586996914 | 1.83917216635334 | -0.45757779580724 |
| H | -3.22496451485969 | 3.98332546470532 | -0.50793140067337 |
| S | -2.53999519989237 | 0.38874767662114 | -0.41676579843495 |
| S | 2.45217762477451 | 4.15832547970478 | -0.49154636964938 |
| C | 2.69973954561933 | 1.56164480800397 | -0.44143968757435 |
| C | 3.45817490108415 | 2.70863067404931 | -0.44861978571382 |
| H | 3.13889035772314 | 0.56363896112613 | -0.40571294046525 |
| C | -4.98607691088532 | 1.71636392812921 | -0.48468860048755 |
| C | 4.89923323959686 | 2.83322456420278 | -0.41747769008923 |
| C | -5.74868415848190 | 0.58595832823325 | -0.75337883540864 |

```
C  -7.13998425848561   0.82425735271978  -0.70194837973719
C  -7.45692909246999   2.13857916965819  -0.38979729804830
C   5.66373240583334   3.98812920667361  -0.53594587073650
C   7.05374297152908   3.75064030672165  -0.46174820370183
C   7.36914806929740   2.41122405356688  -0.28388407791881
S  -6.01248386892212   3.09072539648776  -0.14941471957391
S   5.92416954840359   1.43272094042182  -0.20560667682547
H  -5.30936434809089  -0.38499607455409  -0.99833595853872
H  -7.89268107139172   0.05555328124668  -0.89313875593319
H   5.22689785849396   4.97880524771707  -0.67989432944009
H   7.80670406129517   4.53889903691741  -0.53689007382553
C  -8.74809771939937   2.76305725508905  -0.24207430969245
N  -8.91700898205671   4.03109819857182   0.07123990495418
S -10.25260978451903   1.82985509135049  -0.47489706221237
C -11.14111783928679   3.28609564282605  -0.11434481140447
C -10.22876702442485   4.31624443081158   0.14113571488590
S -11.11554142203778   5.77087905912586   0.50444203545076
N -12.45160325146580   3.57226659459540  -0.04058756685992
C -12.62015240159998   4.83879626063649   0.27655283492593
C   8.65885262990604   1.77994495342166  -0.15622207231327
S  10.16264465250065   2.74086026634125  -0.22541130099115
N   8.82753439164649   0.48583622711076   0.02175872171866
C  11.04997512091656   1.25823952694831   0.00735869968471
C  10.13820764221044   0.20148042055896   0.11193978238672
N  12.35910291078117   0.97287545948547   0.10262086336069
S  11.02415632291100  -1.27935941245539   0.35059446038939
C  12.52742226134187  -0.31984470686023   0.28699500505563
C  13.81847557322089  -0.94533791230100   0.43654096212125
C  14.13396908183654  -2.28210752615110   0.65781272084544
C  15.51903157840738  -2.50973353854830   0.76030723471160
C  16.28646447109758  -1.35295869405360   0.61844801439235
S  15.25561211019501   0.03818064351927   0.35345187688764
H  13.37962920650839  -3.06745846368719   0.74615887815887
H  15.96706906659122  -3.49026843943801   0.93762924687983
C -13.91215484552326   5.45932071872376   0.43877007128120
C -14.22600308802425   6.76921012817092   0.78673881101545
C -15.61228821739567   7.00308974476589   0.85131596789516
C -16.38209190420925   5.87770961214653   0.55430494129288
S -15.35215798960305   4.50887549324638   0.18920653475322
H -16.05958159454388   7.96551492966097   1.11053405376079
H -13.46972411809614   7.53086596319901   0.99043143818597
C -17.81415113777683   5.83854005495503   0.54919869653700
C -18.67199844155407   4.81498224540758   0.28549509990756
H -18.31091067596430   6.78412900023321   0.79954312806392
S -18.31218387456061   3.14130495630601  -0.15140957238790
C -20.03990367154395   2.71247660006809  -0.25067217235130
S -20.60604075070600   1.22498634779449  -0.63193331993382
N -20.78586821868358   3.82482457251039   0.04282900983134
C -20.14382470691453   5.02992582706189   0.35036276206651
H -21.80199311647406   3.77153593876931   0.03637157716732
O -20.72333349857732   6.05794021659887   0.62040375041090
C  17.71654022144448  -1.30048933749267   0.68472373946139
C  18.57095876047997  -0.24750587848534   0.56479869290134
H  18.21446714686427  -2.26144646104969   0.86402195328915
S  18.20706125259060   1.45703126460411   0.27624876148729
C  19.93029801293397   1.91213526075779   0.30664983078729
C  20.04099197382846  -0.45218690673676   0.68152309996662
O  20.62271179088529  -1.49552797329221   0.87784139836668
```

| | | | |
|---|---|---|---|
| N | 20.67841076958747 | 0.78439457145148 | 0.52674464093345 |
| H | 21.69251674556168 | 0.84939374294627 | 0.57921336808857 |
| S | 20.49048765937453 | 3.43630073962386 | 0.10204848296610 |
| C | -0.56505841419629 | 5.15908391910894 | -0.48663237252296 |
| C | -0.04010057802059 | 6.10722141319768 | -1.34364151841719 |
| C | -0.45881815967383 | 7.43342795679724 | -1.05547204215768 |
| C | -1.30476144834381 | 7.52015128052422 | 0.03565334472838 |
| H | -0.16672481963933 | 8.30225281216378 | -1.64958853311934 |
| H | 0.60976946446892 | 5.84814929149044 | -2.18214339721599 |
| S | -1.56995047160452 | 5.93447591863290 | 0.72198166965807 |
| C | 0.47898784656745 | -0.61318908612903 | -0.43354868158552 |
| C | -0.05268598746699 | -1.56623031956739 | 0.41390133235421 |
| C | 0.36814875281050 | -2.89084410838778 | 0.12156911411580 |
| C | 1.22260439551993 | -2.97165262983646 | -0.96333643739934 |
| H | 0.07115492704126 | -3.76278834575594 | 0.70861591342068 |
| H | -0.70890227366933 | -1.31233882529408 | 1.24904800523914 |
| S | 1.49352602144833 | -1.38222029708981 | -1.63864111699218 |
| C | 1.85333778108371 | -4.14106953373074 | -1.54940057738336 |
| C | 2.95348025843991 | -4.21054539518207 | -2.38601407082948 |
| C | 3.29269622105195 | -5.54138866468178 | -2.76338089556503 |
| C | 2.45264818724999 | -6.47830235659186 | -2.21372915001197 |
| S | 1.22753040908849 | -5.74653850961638 | -1.23744120744397 |
| H | 4.13394551943065 | -5.79434332808951 | -3.41321245966820 |
| H | 2.48053736076227 | -7.56259292416587 | -2.33014186000414 |
| H | 3.51176423231136 | -3.32752259096982 | -2.70543594866620 |
| C | -1.93115090287524 | 8.69272400433025 | 0.62010242522414 |
| C | -3.02660801184285 | 8.76690201446271 | 1.46241975202361 |
| C | -3.36248911240117 | 10.09975062039754 | 1.83568529432471 |
| C | -2.52457305127911 | 11.03343258923099 | 1.27734212035293 |
| S | -1.30542139858846 | 10.29615053426398 | 0.29778907271758 |
| H | -3.58406407833500 | 7.88586355661956 | 1.78869006080705 |
| H | -4.19995711822286 | 10.35639150957923 | 2.48896280519713 |
| H | -2.55084693174215 | 12.11826351401434 | 1.38907560783680 |

Optimized geometry of BTTzRv3
| | | | |
|---|---|---|---|
| C | -1.10123745974048 | 1.38289977344552 | -0.41005895622014 |
| C | 0.20045548020468 | 0.85975120141104 | -0.45119880356502 |
| C | 1.25836215897219 | 1.80959348418930 | -0.46239693987445 |
| C | 0.96682673431950 | 3.20962538009953 | -0.44937868510014 |
| C | -0.33462028763983 | 3.73265529356208 | -0.40748957064948 |
| C | -1.39244706240963 | 2.78289122933531 | -0.38913725087179 |
| C | -2.80472130523028 | 3.02909760131425 | -0.41596405233383 |
| C | -3.57169671796500 | 1.88847039005729 | -0.44414890476117 |
| H | -3.23649795629826 | 4.03057822041045 | -0.44043455506058 |
| S | -2.57800683656342 | 0.43032922841262 | -0.43854096580389 |
| S | 2.44253058237656 | 4.16304757662551 | -0.41591814014913 |
| C | 2.67061522737145 | 1.56448981132284 | -0.43222159985366 |
| C | 3.43763764616337 | 2.70552307902761 | -0.41163068752945 |
| H | 3.10264659822640 | 0.56279474279841 | -0.42387097496434 |
| C | -5.01418010276764 | 1.77750215825606 | -0.47752009723443 |
| C | 4.87972038630412 | 2.81840125844880 | -0.38131259639175 |
| C | -5.78476479939722 | 0.65729629959420 | -0.76560426670745 |
| C | -7.17430466398982 | 0.90597259312953 | -0.71625172253060 |
| C | -7.48195614455282 | 2.21818753903529 | -0.38646301184907 |
| C | 5.65353752636891 | 3.96760021326839 | -0.49396480175942 |
| C | 7.04173942533303 | 3.71775484857169 | -0.42575923166562 |
| C | 7.34618980776468 | 2.37451564694596 | -0.25793727492261 |
| S | -6.03102716411936 | 3.15519535880690 | -0.12676455123246 |
| S | 5.89317119695979 | 1.40797983346796 | -0.18050493473335 |
| H | -5.35238895950480 | -0.31217349416416 | -1.02254939019281 |
| H | -7.93235789362035 | 0.14628064241433 | -0.92168677131575 |
| H | 5.22462679507940 | 4.96264460995923 | -0.63135873531761 |
| H | 7.80126018465398 | 4.49990409406708 | -0.49868999972165 |
| C | -8.76853721531932 | 2.85147414380970 | -0.23622958787683 |
| N | -8.92799668506092 | 4.11609722397016 | 0.09509052645693 |
| S | -10.27979616886436 | 1.93571786326221 | -0.49281455705294 |
| C | -11.15743916598478 | 3.39446090686468 | -0.11592601753894 |
| C | -10.23747429359427 | 4.41224951620235 | 0.16090740259753 |
| S | -11.11322823968433 | 5.86962122788977 | 0.53954825833857 |
| N | -12.46573164942597 | 3.69179575526179 | -0.04668796327577 |
| C | -12.62458911185850 | 4.95503943252135 | 0.28809924365921 |
| C | 8.63081554216527 | 1.73099813633599 | -0.14017449110410 |
| S | 10.14290704934930 | 2.67858666217700 | -0.20994927603499 |
| N | 8.78847510951178 | 0.43422870890740 | 0.02779355516184 |
| C | 11.01775979524275 | 1.18613762056516 | 0.00699091509418 |
| C | 10.09686885836667 | 0.13711962010966 | 0.10904700364375 |
| N | 12.32472476754887 | 0.88779901592193 | 0.09226853347428 |
| S | 10.97022712418772 | -1.35364364564422 | 0.33171233332764 |
| C | 12.48187551721923 | -0.40781719600456 | 0.26593585735243 |
| C | 13.76784761792757 | -1.04654255998398 | 0.40248198605914 |
| C | 14.07192995611569 | -2.38777019111876 | 0.61238059852272 |
| C | 15.45541341727522 | -2.62946253592237 | 0.70334353938897 |
| C | 16.23300351957520 | -1.47922642637334 | 0.56370675978881 |
| S | 15.21380135876112 | -0.07636292639221 | 0.31584983294997 |
| H | 13.31062704149489 | -3.16636752641170 | 0.70077506916829 |
| H | 15.89520078215862 | -3.61546716811557 | 0.87062341179321 |
| C | -13.91177708670707 | 5.58557612951650 | 0.44979127260173 |
| C | -14.21550269919417 | 6.89276544832044 | 0.81631010393584 |
| C | -15.59987342091494 | 7.13983786911838 | 0.87213337354061 |
| C | -16.37828459324867 | 6.02742987223711 | 0.54977037701555 |
| S | -15.35897201627329 | 4.65411538649862 | 0.17187648519490 |
| H | -16.03975146783846 | 8.10256977669848 | 1.14258798410754 |
| H | -13.45340153599536 | 7.64317071404799 | 1.03909903704793 |
| C | -17.81054064399479 | 6.00351339802976 | 0.53002870828401 |
| C | -18.67637804601675 | 4.99374505989417 | 0.24054696323551 |
| H | -18.29989599150857 | 6.95008993493994 | 0.79103031015476 |
| S | -18.32997613523794 | 3.32386501568794 | -0.22092332810821 |
| C | -20.06113567088445 | 2.91590311721643 | -0.34618104423606 |
| S | -20.63919593521774 | 1.44141316214760 | -0.75883394536365 |
| N | -20.79808353470679 | 4.03125608763357 | -0.04180923384943 |
| C | -20.14637261877223 | 5.22375424659081 | 0.29331682508478 |
| H | -21.81459404048573 | 3.98962129725319 | -0.06052758246500 |
| O | -20.71771780758656 | 6.25331143877802 | 0.57467906579389 |
| C | 17.66397098208305 | -1.44107450280242 | 0.61951930129295 |
| C | 18.52770147861063 | -0.39576352440683 | 0.49930379203179 |
| H | 18.15392816902238 | -2.40788828505620 | 0.78914549761730 |
| S | 18.17846955129318 | 1.31390285861660 | 0.22345208683110 |
| C | 19.90639684851708 | 1.75168412371341 | 0.24218904663085 |
| C | 19.99649703993867 | -0.61563443637848 | 0.60297130824739 |
| O | 20.56946660795392 | -1.66574271034214 | 0.78873182381320 |
| N | 20.64494817486765 | 0.61540449824134 | 0.44986556035689 |
| H | 21.66006203604224 | 0.66982302759571 | 0.49421717873361 |
| S | 20.48000370731436 | 3.27141285793363 | 0.04154031006818 |
| C | -0.56748615556547 | 5.18568782873257 | -0.38821086760044 |

```
C  -0.02770845088942    6.15158640971498   -1.21636000344986
C  -0.43735959242444    7.47267286425487   -0.89849118766896
C  -1.29310765887905    7.54004796481347    0.18824636616376
H  -0.12923070516547    8.35314109149479   -1.46671793077973
H   0.62795828906703    5.90908132100831   -2.05526360572526
S  -1.57931324550366    5.93798119173878    0.83001944802656
C   0.43320568397551   -0.59314043524264   -0.47472234887104
C  -0.11221534476294   -1.56243226688752    0.34577357582806
C   0.29882054147416   -2.88233923207578    0.02476947084979
C   1.16142624466363   -2.94553629421109   -1.05674953674136
H  -0.01342074706419   -3.76499712839035    0.58730595160860
H  -0.77293241647071   -1.32355678067310    1.18176086484774
S   1.45260414749534   -1.34084043361602   -1.68977109825131
C   1.78459910035371   -4.10451552311756   -1.66177725915243
C   2.81478943876300   -4.16094956530123   -2.58571824893965
C   3.16909933215915   -5.48056671201620   -2.96106056359148
C   2.41889742068727   -6.45833463218915   -2.32955911965101
S   1.25239014973643   -5.72223152953540   -1.25202025584322
H   3.95361088240239   -5.71191572691985   -3.68575756821166
H   3.30836260307343   -3.26944345432521   -2.97887045334752
C  -1.91296940034816    8.70130320901663    0.79233521072278
C  -2.93816445399174    8.76117746506276    1.72159128331319
C  -3.29019203653184   10.08219326803800    2.09413867011977
C  -2.54333051787773   11.05758436363049    1.45503789845514
S  -1.38284091496004   10.31743737299071    0.37382777116935
H  -3.42975010697718    7.87116960053370    2.12058535511946
H  -4.07061702810825   10.31625760590751    2.82237299875864
C  -2.62079503357969   12.50016905624032    1.58164560183237
C  -1.73147285463178   13.46332761205121    1.13625962100327
C  -2.13549378005633   14.79324931460816    1.44621942481357
C  -3.32778236902849   14.83996179599260    2.12574329880647
H  -0.80404726913526   13.21735808079875    0.61353985044391
H  -1.55939818090227   15.68188227820881    1.17830077828812
S  -3.97840066523442   13.25924027190765    2.38839396625607
H  -3.86567415360706   15.71617668627056    2.48878679677517
C   2.49734433431565   -7.90045071651930   -2.46081927947250
C   1.60617972570135   -8.86534803987005   -2.02291003365821
C   2.01187874996736  -10.19408061587136   -2.33575410217194
C   3.20717339206711  -10.23810751187573   -3.01017712737227
H   0.67630396708421   -8.62139071144277   -1.50362354162866
H   1.43467465667123  -11.08377127654508   -2.07377346274890
S   3.85908906208345   -8.65640844499774   -3.26347782651840
H   3.74680668548786  -11.11289575476165   -3.37408552098615
```

Optimized geometry of BTTzRv3 (cation)
```
C  -1.10486140120375    1.36502890064997   -0.07760197356317
C   0.19826194549504    0.84224283419434   -0.14382958722662
C   1.26183365322334    1.80816725629625   -0.14875957531411
C   0.96636581168399    3.20590011913141   -0.10325949836085
C  -0.33668443149854    3.72857339128837   -0.03594120862842
C  -1.40019787403747    2.76268799753752   -0.03247004316050
C  -2.80078274696296    3.00867333706246   -0.08437380539456
C  -3.57686212231048    1.86293766216924   -0.14772357598160
H  -3.23731455231455    4.00739783752249   -0.11337119472875
S  -2.58108594705975    0.41558718688887   -0.15856736713198
S   2.44243365763920    4.15565920399198   -0.02555084250087
```

```
C   2.66265488128406    1.56266319377503   -0.10113075850750
C   3.43875187311212    2.70865171407985   -0.04164438725167
H   3.09958539515388    0.56406588640384   -0.07423240725236
C  -5.01060219185492    1.76461692231376   -0.21342304535961
C   4.87275071913592    2.80814076199971    0.01278947790039
C  -5.78634609174096    0.62431679228373   -0.41730338982514
C  -7.16889445597576    0.88522736087276   -0.43924920539810
C  -7.47323104731850    2.23048182797376   -0.24842034837679
C   5.64892177088878    3.95057024744414    0.20304025196874
C   7.03201000365373    3.69238011119436    0.20913766613986
C   7.33658737395814    2.34685867320192    0.02007762067648
S  -6.02607107703156    3.18024511918183   -0.03632848015534
S   5.88868477747599    1.39317541441693   -0.16774842699878
H  -5.35880112487665   -0.37109331612417   -0.55793868469215
H  -7.92852670639071    0.11550813481730   -0.59318951131745
H   5.22137165811965    4.94595118960172    0.34374926558849
H   7.79201820717766    4.46421630843128    0.35008261434159
C  -8.75548852962206    2.88184530889003   -0.20585514834258
N  -8.90821296802832    4.17616974918152    0.00032613818261
S -10.26664799000828    1.96360151458515   -0.43601026600821
C -11.13458315979755    3.45735297727982   -0.23265295127155
C -10.21081317751311    4.49218609313454   -0.01382486744364
S -11.08304085412028    5.98663836242088    0.18936761451615
N -12.43798933158582    3.76922996993348   -0.24653810122821
C -12.59544403169138    5.06206259779728   -0.03847906043714
C   8.61989782793493    1.69911306590527   -0.04010332422921
S  10.13214403447228    2.62681839121861    0.14072453476494
N   8.77347238647966    0.40273424526468   -0.23229310486448
C  11.00154651950808    1.13396358380465   -0.06237341096939
C  10.07744256384691    0.09236025363260   -0.24495365751021
N  12.30631446380863    0.82813494758625   -0.07733412767964
S  10.95153518379803   -1.40061293674901   -0.45161037152366
C  12.46475635690284   -0.46648776105747   -0.27324823726501
C  13.74882144963929   -1.11034881885503   -0.34639788687009
C  14.05558304229747   -2.45165386811097   -0.56303198725709
C  15.43837451930144   -2.70261154320841   -0.57416628703302
C  16.21361910778420   -1.55891549986631   -0.36591917078939
S  15.19465628057952   -0.15300003362732   -0.15552420712893
H  13.29633492666681   -3.22336495314959   -0.71084230868503
H  15.88130791277009   -3.68853849533073   -0.73160581569066
C -13.87759475655997    5.71231606228482    0.00702044716296
C -14.18277006812098    7.05168568063879    0.23733110699253
C -15.56353228927291    7.31208067955309    0.20732374592322
C -16.33865789181540    6.17769301678902   -0.04706260766942
S -15.32232444891953    4.76876643563397   -0.24946407450846
H -16.00519675592493    8.29807560579785    0.36777302478577
H -13.42395358197509    7.81518668980583    0.42462785162969
C -17.76943864377158    6.16068780995150   -0.13206895307351
C -18.62464598818788    5.12946301121005   -0.37224533187539
H -18.26368076681946    7.12829744351558    0.01916183036431
S -18.27007464503617    3.42872666858608   -0.66440128533564
C -20.00039693723539    3.01531917352940   -0.84232865697165
S -20.55868636752009    1.51212828035492   -1.15302380198451
N -20.74225570721837    4.15632355743787   -0.67513069535044
C -20.09821499499301    5.36976271108855   -0.41428084049366
H -21.75716202616850    4.11848047959249   -0.74447024200007
O -20.66642113122144    6.42455649648091   -0.25406762349687
C  17.64629536903103   -1.53012988499918   -0.33194788875995
```

```
C  18.50108553931957   -0.48831674070163   -0.14082315759281
H  18.14282533543354   -2.49609279321980   -0.48566547798514
S  18.14279489331048    1.21376590433700    0.13778776573840
C  19.87465741910890    1.64537912046603    0.23776898860975
C  19.97735012016303   -0.71515504940956   -0.15404958741502
O  20.54850852341408   -1.76756790873819   -0.31936917701991
N  20.61995998567497    0.50826935629907    0.05957939258773
H  21.63651078453862    0.55667799238548    0.08456179093283
S  20.43064179459875    3.15867253904833    0.49997745426720
C  -0.57365220868679    5.16643114054694    0.01385209776763
C   0.07880117755708    6.16584030237687   -0.70229708925565
C  -0.35958267212353    7.46900552662736   -0.39631681068629
C  -1.35339091957608    7.50800878418817    0.57981833547939
H   0.02591708004934    8.36535979760620   -0.88632909198684
H   0.82746958185263    5.95136713000157   -1.46720626145103
S  -1.73890816808160    5.88739679615893    1.11294417887892
C   0.43428498719470   -0.59584079295250   -0.19113389085634
C  -0.21996661091485   -1.59320607842463    0.52625612978115
C   0.21685596702424   -2.89747486811690    0.22281510376277
C   1.21103820023621   -2.93950623350239   -0.75281616272380
H  -0.17026283649309   -3.79247042296265    0.71403149523334
H  -0.96903378813047   -1.37640303166355    1.29013848380706
S   1.59919192976266   -1.32031954356221   -1.28834661168321
C   1.87420742845998   -4.08545514791642   -1.31198084792127
C   2.92365799144643   -4.13107012329502   -2.22560937429253
C   3.32642659419894   -5.43883037259878   -2.55826860182654
C   2.59616253639358   -6.43044030886551   -1.90782506672969
S   1.38686325514857   -5.71101476665425   -0.87054553037366
H   4.13678235536349   -5.66214015086391   -3.25600324954303
H   3.39144595325877   -3.23438825286382   -2.63815387467680
C  -2.01800636047322    8.65207133482127    1.14114746919533
C  -3.06732351788902    8.69464567748517    2.05506564418449
C  -3.47179833177040   10.00126629501347    2.39012862607936
C  -2.74301566935351   10.99501932163968    1.74130723201438
S  -1.53295523326248   10.27907160741022    0.70250183579107
H  -3.53380297717743    7.79660353285362    2.46612860881799
H  -4.28233855605535   10.22221580860910    3.08839954167668
C  -2.88171776887543   12.42823686908560    1.83810120636647
C  -2.16309830763435   13.41902676255144    1.18238715496288
C  -2.57487863787162   14.73329831539293    1.52524074077424
C  -3.60402600046479   14.74273393014835    2.43875845447279
H  -1.36048034166864   13.20204901493392    0.47277019883235
H  -2.12818074865607   15.64000210397076    1.11235352915402
S  -4.08207282212689   13.14891220169220    2.89291397533100
H  -4.11039799998655   15.60490832491793    2.87410134589740
C   2.73293442934313   -7.86401541299347   -2.00204761513577
C   2.01235635917058   -8.85266624367935   -1.34525357099665
C   2.42267301486001  -10.16809926896695   -1.68540316332346
C   3.45270471479253  -10.18054425471346   -2.59788682001930
H   1.20942266005991   -8.63334927686539   -0.63671425116501
H   1.97442924080647  -11.07346435854653   -1.27125562759776
S   3.93301776743470   -8.58818475274093   -3.05477103901529
H   3.95836799817754  -11.04417847228224   -3.03115440918849
```